\documentstyle[emlines2]{article}
\def\be{\begin{eqnarray}}
\def\ee{\end{eqnarray}}
\def\nn{\nonumber}

\begin{document}

\hfill ITEP-TH-56/98 \\
\phantom. \hfill hepth/9810031 \\

\centerline{\Large{Identities between Quantum Field Theories}}
\centerline{\Large{in Different Dimensions}}

\bigskip

\centerline{A.Morozov}

\centerline{117259, ITEP, Moscow, Russia}

\bigskip

\centerline{Talk given at INTAS-RFBR School at Como, Italy,
September 1998}

\bigskip

\centerline{ABSTRACT}

\bigskip

Review of some old and relatively new ideas surrounding
the subject of AdS/CFT correspondence, generalized tau-functions
and possible equivalences between a priori different quantum
field theories.

\bigskip

\bigskip


This talk is motivated by the recent discussions around
the subject of the ``ADS/CFT correspondence'', discovered
in \cite{Mal} and reformulated in \cite{GKP}, \cite{WAdS}.
In \cite{WAdS} the issue was actually reduced to the problem of
various representations of generalized $\tau$-functions, which has
been encountered in various other contexts during last years.
Particular subject of AdS/CFT correspondence emphasizes the
possibility to represent one an the same effective action
in terms of quantum field theory models in different dimensions,
and it is this aspect of $\tau$-function theory that will be
briefly reviewed in the present notes.

{\bf 1.} The AdS/CFT correspondence itself was discovered
in the context of brane theory.
The simplest view on branes is in the framework
of ``tomography approach'' (generalized Radon transform) \cite{tom}.
One can study any given (quantum) theory,
e.g. the entire ``Theory of Everything'' in its $D$-dimensional
($D$=10 or 11) phase, by looking at the
propagation of probe objects in the background fields.
Then some basic properties of entire theory,
i.e. the properties of background fields, can be recovered from
the behavior peculiarities of the probe objects.

If probe objects are particles (0-branes), this idea is
realized in conventional tomography devices, used in modern
medicine.
Inverse Radon transformation of this type is somewhat
complicated: to recover the pattern of the body one needs to
collect information by sending rays from all possible
directions and examine them all together.
One can instead make use of the quantum properties of
particles, namely their wave properties leading to
interference and diffraction, and obtain a holography
picture in codimension one (on a ``screen'')
still carrying complete information about the
full multidimensional structure in the bulk.
Inverse holography is practically simple (it is enough
to shed the light on holography plate) but formally it
is still a sophisticated transformation.
In any case tomography with the help of particles does
not translate description of original
multidimensional theory into a pure particle ($d=1+0$)
problem: additional structures like screens of
codimension one are always needed.
Moreover, if we switch from technological applications
to the fundamental theories and
fundamental ``laws of nature'', they do not seem to turn into
anything reasonably understandable under such particle-tomography
transformation.

It is well known that if the probe particles are
substituted by probe ``relativistic'' strings
(1-dimensional objects with constant tension),
quantum tomography becomes very efficient
exactly in application to the fundamental theories.
Namely, the fundamental equations of motion of
original bulk theory (like Einstein equation
$R_{\mu\nu} = 0$ etc) turn into a {\it symmetry
principle} for the quantum theory of probe strings:
effective $d=2$ theory on the string world sheet
becomes conformal invariant (under an infinite
symmetry of local Weyl transformations of
the world sheet metric $g_{ab}(z) \rightarrow
\lambda(z)g_{ab}(z)$). This symmetry ensures decoupling
of non-unitary excitations in effective theory on
the world sheet, which would be associated with
instability of the probe string in {\it of-shell} external
fields of the bulk theory.
In other words, the fundamental laws (equations
of motion of the bulk theory) are the necessary conditions
for stability of probe strings. This observation implies
that strings (1-branes) are in fact among the true excitations
(quasiparticles) of the bulk theory. Of course this
does not exclude particles (the 0-branes) as other possible
quasiparticles: it is just more sophisticated task to
formulate the conditions of their stability,
associated symmetry principle is space-time gauge invariance
and it is not (yet?) reduced to any simple property of
the world-line effective theory.
In other words, no simple way is known
to formulate gauge symmetry in terms of the first-quantized
particles, but when particles are substituted by strings
gauge invariance becomes related to
the requirement of $d=2$ conformal invariance.

An old natural question is what happens when probe strings
(1-branes) are further substituted by relativistic membranes
(2-branes with constant tension) and higher-dimensional
$p$-branes. For some time, because of concentration on the
beautiful studies of strings, this subject was not in the
center of investigation and only recently it attracted new
attention, when it became widely recognized that stringy
objects can not represent the full set of quasiparticles
of the Theory of Everything in all its possible phases.
Unfortunately, approaches developed in application to
strings are not quite sufficient to attack the problems of
higher-dimensional branes (as ordinary field theory technique
developed for the study of particles is not quite sufficient
for exhaustive description of strings).

The key element of the string theory is the possibility to
switch from the Nambu-Goto quantum measure
\be
e^{-Area} = \exp \left(-\int\sqrt{\det_{ab}
\left[G_{\mu\nu}(x)\partial_ax^\mu\partial_bx^\nu\right]}
d^2z \right)
\label{strnaive}
\ee
in summation
over string $d=2$ world sheets embedded into D-dimensional
space-time to the Polyakov measure like
\be
\exp \left(-\int
\left[G_{\mu\nu}(x)\partial_ax^\mu\partial_bx^\nu\right]
g^{ab}\sqrt g d^2z\right)
\label{strtrue}
\ee
in summation over embeddings {\it and} over world-sheet geometries
$g_{ab}$ (including world-sheet topologies).
This is not a formal transformation (like it is in the case
of particles), but rather a physical principle \cite{Pol}, which allows
one to introduce a relevant quantum object which can really play the
role of a stable quasiparticle -- at least in the adequate
(on-shell) background (bulk) geometries $G_{\mu\nu}(x)$.

Such redefinition is even more important for membranes
and generic branes. The stringy measure $e^{-Area}$ reflects
the fact that the string energy is proportional to its length.
Similarly, for membrane the naive measure $e^{-Volume}$
would follow from the fact that membrane energy is proportional
to its area. However, {\it such} membrane can not exist as
a stable quantum object: unlike string, membrane can be
{\it strongly} deformed without changing its area
(a very high but narrow pick can have tiny {\it area}) and
such {\it strong} fluctuations can not be damped by the naive
measure. Thus an analogue of the {physical} substitute
(\ref{strnaive}) $\rightarrow$ (\ref{strtrue}) is even more
important for generic branes than it was for strings.

Unfortunately, the brane analogue of the measure (\ref{strtrue}),
\be
\mu_p\{{\cal A}\} =
\exp \left(-\sum_I {\cal A}_I \int {\cal O}_I(\Phi(z))
d^{p+1}z\right)
\label{branetrue}
\ee
is not yet discovered and there is no available
way for deductive presentation of the theory of probe branes.
One can instead try various guesses and use them in the
search for the proper principles dictating the choice of the set of
the fields $\Phi(z)$ on the brane world-volume, and the
vertex operators ${\cal O}_I(\Phi(z))$, associated with
particular background fields ${\cal A}_I$ of the main
theory in the bulk.

The first immediate guess is that effective world-volume
theory is much simpler in particularly adjusted backgrounds
than in generic circumstances. For example, the
string model is considerably simplified if
$G_{\mu\nu}(x)$ in (\ref{strtrue}) is not just {\it some} solution
to Einstein equations $R_{\mu\nu} = 0$, but if it is, say,
the {\it flat} solution, $G_{\mu\nu}(x) = \delta_{\mu\nu}$.
Then quantum theory with the weight (\ref{strtrue}) becomes
essentially a theory of free $d=2$ fields $x^\mu(z)$ and all
the correlators in this theory can be obtained by application
of Wick theorem. One can imagine that the same should be
true for generic branes: in  certain backgrounds the brane
model is drastically simplified and it can even happen that the
relevant quantum theory becomes that of the free fields and some
sort of Wick theorem is applicable. The most straightforward way
to find such distinguished backgrounds would be to look at the
back reaction of probe brane on the bulk theory. The probe
object is of course a source of the fields in the bulk, but
normally ``probe'' means that these emitted fields are neglected.
Still the shape of emitted fields can be looked at in order
to determine distinguished backgrounds: background of its own
fields is normally the one in which the object feels ``most
comfortable''. This is a kind of the necessary condition
for the self-consistency of a quasiparticle definition:
it (quasiparticle) can be destabilized by arbitrarily
imposed external fields, but it should not be self-destroyable
by its own fields. Of course, this is not a rigorously formulated
statement, but this argument provides a possible direction for
the search of distinguished backgrounds, presumably preferred by
the brane theory. The $AdS_{p+2}\times S_{D-p-2}$
backgrounds for supersymmetric branes are exactly of this type
(one should take into account that supersymmetric Dirichlet branes
are not only gravitating but also charged w.r.to various gauge fields,
since charges are integer-valued, gravitation effects can not
be negligible, therefore preferred backgrounds are not just flat,
but rather maximally symmetric non-trivial geometries, composed
of spheres and AdS spaces).

The second immediate guess is that the scalar free fields are
not the most natural fields beyond two space-time dimensions,
and effective world-volume theories in $d=p+1$ dimensions should
rather contain $\left[\frac{p+1}{2}\right]$-forms (i.e.
vector fields for 3-branes, 2-forms for 5-branes etc).
If $k$-forms are gauge fields, then for $k\geq 2$ they are
naturally abelian and therefore free. However, for $k=1$
at least the non-abelian gauge symmetry can also occur
(what is the proper language to describe non-abelian 2-forms,
if any, is still unclear). Remarkably, non-abelian Yang-Mills
fields (gauge 1-forms) become essentially free (i.e. the
Wick theorem holds) not only when the gauge group $U(N)$ is small,
$N=1$, but also in the opposite limit $N=\infty$.

{\bf 2.} This follows from the old t'Hooft's calculus \cite{tH}.
The Green functions of bosonic fields like $A^a_\mu$ are
basically given by the Coulomb law:
\be
\langle A^a_\mu(z_1) A^b_\nu(z_2) \rangle \sim
\left(\delta_{\mu\nu} -
\frac{\partial_\mu\partial_\nu}{\partial^2}\right)
\frac{\delta^{ab}}{|z_1 - z_2|^{d-2}}
\label{Aprop}
\ee
However, $A^a_\mu(z)$ is not the relevant operator in
Yang-Mills theory. The simplest gauge invariant operator is
rather ${\rm Tr} F_{\mu\nu}^2(z)$ with
$F_{\mu\nu} = \partial_\mu A_\nu - \partial_\nu A_\mu +
[A_\mu,A_\nu]$.
This operator contains three different contributions:
quadratic, cubic and quartic in $A$.
Accordingly, there are three contributions to
the pair correlators of such operators which differ
not only in their dependencies on the world-volume
coordinates $z$, but also in those on the
coupling constant $g^2$ and the size $N$ of the gauge group.
The leading term (with the minimal number of $A$-loops) in
quadratic contribution is
\be
\sim \frac{g^4 N^2}{|z_1 - z_2|^{2d}},
\label{num1}
\ee
that in the cubic one is
\be
\sim \frac{g^6 N^3}{|z_1 - z_2|^{3d-4}}
\label{num2}
\ee
and that in the quartic one is
\be
\sim \frac{g^8 N^4}{|z_1 - z_2|^{4(d-2)}}
\label{num3}
\ee
The $g^2$ and $N$ dependencies of various contributions
to the multipoint correlator
$\prod_{i=1}^{2n} {\rm Tr} F_{\mu\nu}^2(z_i)$ are
evaluated with the help of Euler theorem, $V-E+L = 2l$:
\be
g^{2E}N^L = N^{2l - V} \prod_k (g^2N)^{kV_k/2}
\label{numer}
\ee
Here $V,E,L$ stand for the number of vertices, edges (links)
and loops in the Feynman diagram, $V = \sum_k V_k$
where $V_k$ is the number of vertices with multiplicity
$k = 2,3,4$ and $l$ denotes the number of connected
components of the diagram.

From (\ref{numer}) it follows that at large $N$ the
maximally disconnected diagrams (with maximal $l$)
are dominant. Since minimal number of operators
${\rm Tr} F_{\mu\nu}^2(z)$
in connected sub-diagram is two, we conclude that
the $2n$-fold correlator in the large $N$ limit is
reduced to the product of $n$ elementary pairwise correlators.
This is already almost the Wick theorem. One needs
only to get rid of sophisticated combinatorial factors
(and the mixture of different $z_i$-dependencies if $d\neq 4$),
which occur if all the three contributions
(\ref{num1}-\ref{num3}) are simultaneously taken into
account. The three contributions are all comparable in
the standard t'Hooft's limit $N\rightarrow\infty$,
with $g^2N$ finite. Further simplifications  arise
(and Wick theorem indeed holds) in two other ``double-scaling''
large-$N$ limits, when $g^2N \rightarrow 0$ (``abelian''
large-$N$ where multiplicities $k$'s should be minimized to $k=2$)
or $g^2N \rightarrow \infty$ (strongly non-abelian
``long wave'' or Maldacena's limit where $k$'s should be maximized
to $k=4$ and all the derivatives of $A$-fields are neglected).
The elementary pairwise correlators in these two cases
are given by (\ref{num1}) and (\ref{num3}) respectively.

{\bf 3.} Occurrence of the Wick theorem implies the possibility
to introduce some new free fields $\varphi(z)$ instead of original
composite operators ${\rm Tr} F_{\mu\nu}^2(z)$.
However, their elementary pairwise correlators --
both (\ref{num1}) and (\ref{num3}) --
are somewhat unusual from
the point of view of the $z$-dependencies: both degrees
$2d$ and $4(d-2)$ (which occasionally coincide for $d=4$,
i.e. exactly when vector fields should be most relevant)
can seem somewhat un-natural for the free fields in $d$
space-time dimensions.
Remarkably, at least one of these degrees, $2d$ possesses
a natural interpretation.
Namely, the Green function of {\it free} scalars in
$d+1$-dimensional space $AdS_{d+1}$ (i.e. the space with the metric
$ds^2 = \frac{t^2 + |z|^2}{t^2}$ and the Ricchi tensor
$g_{mn} = \Lambda g_{mn}$) is equal to
\be
\langle \varphi(t,z_1) \varphi(0,z_2) \rangle \sim
\frac{t^{d+1}}{(t^2 + |z_1 - z_2|^2)^d}
\label{adsgf}
\ee
which in the limit $t\rightarrow 0$, i.e. on the $d$-dimensional
boundary of $AdS_{d+1} = AdS_{p+2}$, turns -- after appropriate
rescaling of fields -- into exactly the relevant formula
$|z_1-z_2|^{-2d}$.

{\bf 4.} Certain evidence in support of ``holography principle''
in quantum gravity comes from considerations of
propagation in {\it classical} gravitational fields
\cite{tHgrav,Sus}.
The basic idea is to note that geodesics are non-straight
lines in the presence of gravitating bodies and combine this
with the fact that all physical objects are in fact gravitating.
When combined, these two observations imply
that {\it projection} from the space on its
boundary (a ``screen'' of codimension one) can have certain
peculiarities when gravity is taken into account.
The main phenomenon is illustrated in Fig.1 \cite{Sus}.
Ordinary projection by the rays orthogonal to the screen
leads to information loss, since objects (point $A$) in the
shadow of object $B$ are not seen.
However, if one takes into account the fact that the object $B$ is
gravitating, the point $A$ can still be seen in {\it orthogonal}
projection, since the rays (of light) are no longer straight.
Certain calculations involving black holes \cite{tHgrav,Sus}
demonstrate that the idea can be formulated in a self-consistent
way and indeed reflects some  essential property of general relativity.

\bigskip

\bigskip

\bigskip

\special{em:linewidth 0.4pt}
\unitlength 0.70mm
\linethickness{0.4pt}
\begin{picture}(145.00,68.67)
\emline{10.00}{20.00}{1}{69.67}{20.00}{2}
\emline{85.00}{20.00}{3}{145.00}{20.00}{4}
\put(115.00,44.67){\circle{10.67}}
\put(40.00,45.00){\circle{10.00}}
\put(40.00,65.00){\circle*{0.67}}
\put(115.00,65.00){\circle*{0.67}}
\emline{35.00}{45.00}{5}{35.00}{20.00}{6}
\emline{45.00}{45.00}{7}{45.00}{20.00}{8}
\emline{115.00}{65.00}{9}{100.33}{20.00}{10}
\emline{115.00}{65.00}{11}{129.00}{20.00}{12}
\emline{115.00}{65.00}{13}{89.33}{20.00}{14}
\emline{115.00}{65.00}{15}{140.67}{20.00}{16}
\emline{10.00}{19.00}{17}{70.00}{19.00}{18}
\emline{85.00}{19.00}{19}{145.00}{19.00}{20}
\put(71.67,12.67){\makebox(0,0)[rb]{screen}}
\emline{40.00}{65.00}{21}{38.60}{64.38}{22}
\emline{38.60}{64.38}{23}{37.26}{63.61}{24}
\emline{37.26}{63.61}{25}{36.00}{62.68}{26}
\emline{36.00}{62.68}{27}{34.80}{61.59}{28}
\emline{34.80}{61.59}{29}{33.68}{60.35}{30}
\emline{33.68}{60.35}{31}{32.62}{58.96}{32}
\emline{32.62}{58.96}{33}{31.63}{57.42}{34}
\emline{31.63}{57.42}{35}{30.71}{55.71}{36}
\emline{30.71}{55.71}{37}{29.86}{53.86}{38}
\emline{29.86}{53.86}{39}{29.07}{51.85}{40}
\emline{29.07}{51.85}{41}{28.36}{49.69}{42}
\emline{28.36}{49.69}{43}{27.71}{47.37}{44}
\emline{27.71}{47.37}{45}{27.14}{44.89}{46}
\emline{27.14}{44.89}{47}{26.63}{42.27}{48}
\emline{26.63}{42.27}{49}{26.19}{39.49}{50}
\emline{26.19}{39.49}{51}{25.82}{36.55}{52}
\emline{25.82}{36.55}{53}{25.52}{33.46}{54}
\emline{25.52}{33.46}{55}{25.29}{30.21}{56}
\emline{25.29}{30.21}{57}{25.12}{26.81}{58}
\emline{25.12}{26.81}{59}{25.03}{23.26}{60}
\emline{25.03}{23.26}{61}{25.00}{20.00}{62}
\emline{40.00}{65.00}{63}{41.40}{64.06}{64}
\emline{41.40}{64.06}{65}{42.73}{62.99}{66}
\emline{42.73}{62.99}{67}{43.99}{61.79}{68}
\emline{43.99}{61.79}{69}{45.18}{60.46}{70}
\emline{45.18}{60.46}{71}{46.30}{59.00}{72}
\emline{46.30}{59.00}{73}{47.36}{57.42}{74}
\emline{47.36}{57.42}{75}{48.35}{55.70}{76}
\emline{48.35}{55.70}{77}{49.27}{53.86}{78}
\emline{49.27}{53.86}{79}{50.12}{51.89}{80}
\emline{50.12}{51.89}{81}{50.90}{49.79}{82}
\emline{50.90}{49.79}{83}{51.62}{47.56}{84}
\emline{51.62}{47.56}{85}{52.26}{45.20}{86}
\emline{52.26}{45.20}{87}{52.84}{42.72}{88}
\emline{52.84}{42.72}{89}{53.35}{40.11}{90}
\emline{53.35}{40.11}{91}{53.79}{37.36}{92}
\emline{53.79}{37.36}{93}{54.17}{34.49}{94}
\emline{54.17}{34.49}{95}{54.47}{31.49}{96}
\emline{54.47}{31.49}{97}{54.71}{28.36}{98}
\emline{54.71}{28.36}{99}{54.88}{25.11}{100}
\emline{54.88}{25.11}{101}{55.00}{20.00}{102}
\put(33.67,15.33){\makebox(0,0)[lb]{shadow}}
\put(40.00,68.33){\makebox(0,0)[cb]{A}}
\put(115.00,68.67){\makebox(0,0)[cb]{A}}
\put(40.00,45.00){\makebox(0,0)[cc]{B}}
\put(115.00,45.00){\makebox(0,0)[cc]{B}}
\put(40.00,5.00){\makebox(0,0)[cc]{Figure 1}}
\put(115.00,5.00){\makebox(0,0)[cc]{Figure 2}}
\put(150.67,12.34){\makebox(0,0)[rb]{screen}}
\end{picture}

\bigskip

Of course, shadows are not really a problem for collecting
information about the bulk on the codimension-one screen.
Shadows are problem for {\it orthogonal} projection, but
not for anything else. For example, if the bodies are
themselves emitting light (in all directions) or image in
the scattered light is considered (when light can be reflected
in all directions) no finite-size object can prevent another
object from being seen somewhere on the infinite screen
(Fig.2). However, such image exists only if one does {\it not}
impose the requirement for rays to be orthogonal to the screen.
This restriction can seem artificial, but without it
one will have problems with separation of overlapping images
of different objects -- and information will be actually lost.
To recover the information one can either rely upon
quantum (interference) properties of the rays, as one
does in laser holography, or stay in the classical framework
but detect not only the image itself, but also the direction
of the ray. It would be even easier to simply fix the
direction, imposing, for example, the restriction of
orthogonality. However, then one returns
to the problem of shadows and to above observation that
gravitational effects allow one to overcome this problem.

As a next step, one can ask what is the mathematical language
adequate for description of these ideas. If one detects the
end-point of the ray on the screen and its direction, one
actually deals with Cauchy problem for propagating particles/rays.
The relevant statement is that -- unless in specifically bad
circumstances -- the bulk picture (solution to Laplace-like
equation) is uniquely specified by the boundary conditions
on the function {\it and} on its first derivatives.
If, however, one imposes orthogonality restriction the story
is somewhat different. The mathematical problem is now
rather that of analytical continuation, and the
solution is uniquely specified by the boundary condition itself.

The main problem with considerations of this section is that
they are essentially classical -- as are most of their modernized
versions and generalizations involving various configurations
of BPS branes. The simplest question to address is what is
the way to describe fluctuations of vacuum configurations,
like propagating gravitons, in the projected picture.
There is hardly any satisfactory answer to this question at this
moment. Clearly, the picture is very well suited for description
of the topological phase of quantum gravity -- which can hopefully
be constructively developed in the close future,-- but what
is the way to apply it to realistic gravity, which -- at least
naively -- is not quite in a topological phase? Our naive
vacuum in gravity spontaneously(?) breaks gauge invariance and
gauge-non-invariant excitations like individual gravitons seem
to be relevant for the usual description of physics. As usual,
the gauge invariant description is available, but looks a little
bit artificial for description of most phenomena of ordinary
physics -- and transition from this gauge invariant (topological)
description to the ordinary one remains somewhat obscure.
Still it should exist, and information is not really lost
in this transition -- as it is not lost in transition from
solutions to Laplace equation in the bulk to the boundary
conditions in Cauchy or analytical-continuation problems.

{\bf 5.} Topological model is defined by a functional integral
which is independent of the metric (normally, of the metric
in the space-time where the quantum field theory is considered).
This is the context, where one naturally assumes that
quantum gravity is topological -- since the result of
integration (averaging) over all metrics is presumably
independent of any individual metric. This is of course
not so obvious if one tries to give any accurate definition
of the integral -- and there are various non-trivialities
even in the simplest case of 2-dimensional gravity, which
is partly understood. Even if one forgets about such problems,
there are technical difficulties which still prevent one
from dealing with most interesting questions about quantum gravity
and its topological nature. For practical purposes one
substitutes the study of gravities (averages over metrics)
by that of a different class of topological theories \cite{top},
which can be more accurately called cohomological models.
It is still an open question, whether the properties of
cohomological models and gravities are similar and what -- if
any -- are the possible differences.

Cohomological theory
is usually considered as reduction of some larger model with
original Hilbert space ${\cal H}$ and a nilpotent operator
$Q$ acting on this space,
\be
Q:\ \ \ {\cal H} \longrightarrow {\cal H}, \nn\\
Q^2 = 0
\ee
Then the Hilbert space of cohomological theory {\it per se}
is the one of cohomologies of $Q$:
\be
h =\ Ker\ Q/Coker\ Q
\ee
Normally if original model is defined as a field theory on
a compact manifold, associated cohomological one has
small (finite-dimensional) Hilbert space $h$ and
has not too many chances to resemble any field ordinary
field theory with infinite-dimensional Hilbert space
(every particle is an infinite collection of oscillators
with different frequencies and every oscillator has infinitely
many states).
However, things change drastically when the space is
non-compact: boundaries usually increase cohomologies and
cohomological theory in non-compact space-time can be
big enough to resemble (or even become equivalent) to
an ordinary model of quantum field theory.

The simplest example of cohomology increase arises when
one makes a puncture on the Riemann surface (complex curve).
Then $\bar\partial$-cohomologies, which were (finite) collections
of holomorphic forms on original surface are enlarged by
inclusion of infinitely many {\it meromorphic} forms
with poles at the puncture. This example has been intensively
exploited in the theory of open strings \cite{open,open1} and
it can serve to illustrate once again the equivalence between
a model at the boundary and in the bulk and the way in which
information about the bulk is contained in the boundary
model. The relevant bulk/boundary relation in open string
theory is provided by analytical continuation.

{\bf 6.} Let us consider a Riemann surface ${\cal C}$ with holes,
$\partial{\cal C} = \Gamma = \sum_{i=1}^n \Gamma_i$.
Then functional integral
\be
{\cal Z}\{\phi_0\} \equiv \int {\cal D}\phi\
\exp\ \left(-\int_{\cal C} \partial\phi \bar\partial\phi \right)
\ee
which in the case of a closed Riemann surfaces
defines the determinant of Laplace operator
(${\cal Z} \sim \sqrt{ \det N_0/|\det \bar\partial|^2 }$),
depends not only on the moduli of ${\cal C}$ but also
on the boundary conditions $\left.\phi\right|_\Gamma = \phi_0$.
Substituting $\phi = \phi_{cl} + \varphi$ with $\phi_{cl}$
being solution to Laplace equation
$\partial\bar\partial\phi_{cl} = 0$
on ${\cal C}$ with boundary condition
$\left.\phi_{cl}\right|_\Gamma = \phi_0$, so that
$\left.\varphi\right|_\Gamma = 0$, one gets:
\be
{\cal Z}\{\phi_0\} = \sqrt{\frac{\det N_0}{\det_- \Delta_0}}
\ \exp \left(-S_{cl}\{\phi_0\}\right)
\ee
where $\det_-\Delta_0$ stands for deter\-mi\-nant of the scalar
Laplace ope\-rator with Di\-rich\-let boundary conditions and
\be
S_{cl}\{\phi_0\} \equiv \int_{\cal C}\partial\phi_{cl}
\bar\partial\phi_{cl} = \oint_\Gamma \phi_0(\partial_n\phi)_0 =
\nn \\ =
\oint_\Gamma \oint_\Gamma \phi_0(x)
\frac{\partial^2 G_D(x,y)}{\partial n_x\partial n_y} \phi_0(y)
\label{1Scl}
\ee
The classical solution
$\phi_{cl}$ is constructed with the help of
the Dirichlet Green function $G_D(x,y)$
of the scalar Laplace operator, which satisfies
$\left.G_D(x,y)\right|_{x\in\Gamma} = 0$,
$\left.G_D(x,y)\right|_{y\in\Gamma} = 0$:
\be
\phi_{cl}(x) = \oint_{y\in\Gamma}
\frac{\partial G_D(x,y)}{\partial n_y}\phi_0(y)
\ee
where $\partial/\partial n_y$ is derivative in the normal
direction to $\Gamma$.
In the simplest case of an upper half-plane with a straight
line as a boundary, the r.h.s. of (\ref{1Scl}) is
\be
\oint_\Gamma \oint_\Gamma \frac{(\phi_0(x)-\phi_0(y))^2}{(x-y)^2}dx dy
\ee
For generic Riemann surfaces with holes explicit expression involves
special functions made from the Prime bi-differential $E(x,y)$
\cite{Fay} -- like the ordinary Green function \cite{freef}
\be
\frac{\partial^2 G_0(x,y)}{\partial x\partial y} =
\langle\partial\varphi(x) \partial\varphi(y)\rangle
\ = \partial_x\partial_y\log E(x,y)
\label{bosgrf}
\ee
The Dirichlet Green function $G_D$ can be constructed with the help
of the {\it double} (image) technique.

Any Riemann surface with boundaries (in fact, also any
non-oriented Riemann surface) can be represented as a factor
${\cal C} = D/Z_2$ of a closed Riemann surface $D$, a double,
over an antiholomorphic $Z_2$ mapping $z \rightarrow *z$.
For example, an annulus is a factor of rectangular torus
with $\tau = it$ over $Z_2$ mapping $z \rightarrow \bar z$.
The stationary points (in this example they lie on the circles
$Im\ z = 0$ and $Im\ z = t/2$) compose the boundary
$\Gamma = \partial {\cal C}$. (If there are no stationary
points, the factor is non-oriented surface, e.g. the same
torus factored over $z \rightarrow \bar z + \frac{1}{2}$
is non-oriented but closed Klein bottle.)
The genus of the double is $g_D = 2g_{\cal C} + n - 1$,
where $n$ is the number of components of the boundary
$\Gamma = \partial{\cal C}$.
Doubles are not generic closed Riemann surfaces, as in above
example the dimension of the moduli space of doubles is twice
smaller than the dimension of entire moduli space of
the genus $g_D$ surfaces.
Green function $G_D(x,y|{\cal C})$
on a surface ${\cal C}$ with boundaries
is immediately obtained from the ordinary one $G_0(x,y|D)$
on its double:
\be
G_D(x,y|{\cal C}) =
G_0(x,y|D) - G_0(*x,y|D) - G_0(x,*y|D) + G_0(*x,*y|D)
\ee
and on the boundary (i.e. for $x = *x$, $y=*y$)
\be
\frac{\partial^2 G_D(x,y|{\cal C})}{\partial n_x\partial n_y} =
4\frac{\partial^2 G_0(x,y|D)}{\partial x\partial y}
\ee
The r.h.s. is given in (\ref{bosgrf}).

The theory of open strings has various applications.
The one relevant for our purposes is application to the
theory of {\it closed} strings, allowing one to express string
measures on various Riemann surfaces in terms of the same
kind of variables -- points of the Universal Grassmannian
which can be used in the role of the universal module space
\cite{ums}. The basic idea is to make a small hole in the closed
Riemann surface. Given Krichever data \cite{kridata}
-- the complex curve
${\cal C}$, a puncture $z_0$ on it and coordinates $z-z_0$
in the vicinity of the puncture -- one can consider the set
of meromorphic functions (or forms) with poles at $z_0$
puncture and look at their expansion near $z_0$.
One can choose the standard basis in the space of such
meromorphic functions (actually the best choice would be
$1/2$-differentials, but we neglect such details here):
\be
f_n(z) = \frac{1}{(z-z_0)^n} + \sum_{m\geq 0} A_{nm}(z-z_0)^m
\ee
Matrices $A_{nm}$ can be used to introduce coordinates on
the universal moduli space, and all the relevant quantities
are of course invariant under the changes of basises and
coordinates (i.e. the moduli space is actually a
universal Grassmannian). In particular determinant of Laplace
operator is given by \cite{open}:
\be
{\rm det}_D \bar\partial \sim \det (1 - AA^\dagger)
\label{detopen}
\ee

This formalism can be considered as representation of the
``bulk theory'' (of free fields on open Riemann surface, i.e.
of $d=2$ fields in non-trivial gravitational backgrounds)
in terms of the ``boundary theory'' of the $d=1$ ``fields''
$A_{mn}$ (with Universal Grassmannian as a target space).
Expressions like (\ref{detopen}) provide the description
of ``quantum effects in the bulk'' (determinant knows about
all the excitations in the bulk, not just about the
``classical configurations'' -- the zero modes)
in terms of the boundary theory. Thus we see that entire
information about the {\it quantum} theory in the bulk
is preserved at the boundary. In multidimensional case the
similar statement would be that gravitons (and other
particles in the bulk) {\it can} be reconstructed from the
data at the boundary.

{\bf 7.} More examples of identities between field theories in
different dimensions are provided by group theory (which
is nowadays understood to be more or less equivalent
to integrability theory). We shall briefly touch two examples
of this kind: coming from geometrical quantization and
from representation theory for non-compact groups.

The basic chain of relations from the theory of geometrical
quantization (Kirillov-Kostant construction)
which will be of interest for us is:
\be
{\rm Tr}_R 1 \ = \
\int{\cal D}g(t) \exp \oint_{S^1} d^{-1}\Omega_R \ =
\label{KK1}
\\ =
\label{KK2}
\int_{\left.\phi\right|_{\partial D} = X_R}
{\cal D}\phi(t,r){\cal DA}(t,r) \ \exp\left(
\int_D {\rm Tr}\ \phi{\cal F} \ + \oint_{\partial D}
{\rm Tr}\ \phi{\cal A}\right)
\ee
At the l.h.s. here stands the dimension $d_R = {\rm Tr}_R 1$
of representation $R$ of a Lie algebra ${\cal G}$.
It can be considered as a quantity in $d=0$ theory (like
matrix model).
The first relation (\ref{KK1}) expresses it in terms of a $d=1$
functional integral over the group elements $g$  on a circle with the
action which is the $d^{-1}$ of Kirillov-Kostant form $\Omega_R$.
For ${\cal G} = SU(N)$, $\Omega_R$ can be explicitly written as
\be
\Omega_R = {\rm Tr} X_R g^{-1}dg \wedge g^{-1} dg,
\ee
and
\be
d^{-1}\Omega_R = -{\rm Tr} X_R g^{-1}dg
\ee
with some {\it constant} matrix $X_R$,
which specifies the representation $R$.
The closed 2-form $\Omega_R$ is degenerate, it becomes non-degenerate
when restricted to an ``orbit'', associated with representation $R$.
In other words, for given $X_R$ the action $d^{-1}\Omega_R$
is in fact independent of some of the variables $g$, i.e. the
$d=1$ theory in (\ref{KK1}) is a sort of a gauge theory.
Note that in such formulation all the dependence on representation $R$
is concentrated in the action, not in boundary conditions.

The second relation (\ref{KK2}) expresses the same quantity in terms
of $d=2$ gauge model. The gauge field ${\cal A}$ is a 1-form
on the disc $D$ with values in adjoint representation of ${\cal G}$,
its curl is a 2-form ${\cal F} = d{\cal A} + {\cal A}^2$.
The field $\phi$ is a scalar in adjoint representation of ${\cal G}$.
The boundary term in the action, $\oint_{\partial D} {\rm Tr}\phi{\cal A}$,
is gauge invariant because $\phi = X_R = const$ on the boundary
$\partial D = S^1$.
Integral over $\phi$ provides a $\delta$-function of ${\cal F}$,
which implies that ${\cal A}$ is a pure gauge:
${\cal A} = g^{-1}dg$.  Now particular representation $R$ is
specified by the boundary condition.

Relations (\ref{KK1})-(\ref{KK2}) are easily generalized
to character formulas (i.e. to represent
${\rm Tr}_R g$ for the group element $g \neq 1$) and
to quantum groups.
An interesting question is if this chain of relations can
be continued  further to higher dimensions,
in the spirit of hierarchy of anomalies \cite{UFN1},
making use of the operators $K$ and $k$, inverse to the nilpotent
external derivative $d$ and the boundary operator $\partial$,
\be
Kd + dK = I, \nn \\
k\partial + \partial k = I
\ee

{\bf 8.} Another way to change the dimension is to apply (\ref{KK2})
{\it per se} to the case of affine (Kac-Moody) algebras. Then
all dimensions are effectively increased by unity and
(\ref{KK2}) turns into relation between $d=2$
Wess-\-Zumino-\-Novikov-\-Witten (WZNW) model and $d=3$
Chern-Simons theory. Indeed, it is well known \cite{AS} that
Kirillov-Kostant 1-form  $d^{-1}\Omega$ in the case of
affine algebra provides exactly the WZNW action.
As to ${\rm Tr}\ \phi{\cal F} = \sum_a \phi^a {\cal F}^a$,
in the case of the current-algebra realization of affine algebra
the index $a$ includes now a continuous  loop parameter $s$,
so that ${\rm Tr}\ \phi{\cal F} \longrightarrow
\oint ds {\rm Tr}\ \phi(s) {\cal F}(s)$.
As to $F^a(s) = \partial_t A^a_r(s) - \partial_r A^a_t(s) +
[A_t, A_r]^a(s)$, the commutator now contains a piece
$kf^{abc}\left(A^b_t(s)\partial_s A^c_r(s) -
A^b_r(s) \partial_s  A^c_t(s)\right)$
proportional to the central charge $k$. It remains to
change the notation $\phi(s) = A_s(s)$ in order to
recognize that $\int {\rm Tr}\ \phi{\cal F}$
in the case of affine algebras is just the Chern-Simons
action on the $d=3$ manifold with coordinates $t,r,s$ and
the boundary at $r=1$. If the $d=2$ WZNW theory is formulated
on a Riemann surface with coordinates $t$ and $s$,
the Chern-Simons theory, associated to it by (\ref{KK2})
is defined on the $d=3$ manifold filling the Riemann surface
(i.e. the surface is its only boundary).

This identity between the WZNW and Chern-Simons is a simple
example of the general ``AdS/CFT correspondence'' \cite{WAdS},
which states that with any Lie algebra with  generators
$J_\alpha$ one can associate a ``topological'' theory of gauge
fields ${\cal A}$ on a non-compact manifold $M$,
\be
\tau\{A\} \equiv
\langle \exp \sum_\alpha A_\alpha J_\alpha \rangle \ =
\left.\int {\cal DA}\ e^{S_{top}\{{\cal A}\}}\right|_
{\left.{\cal A}\right|_{\partial M} = \{A_\alpha\}}
\label{basrel}
\ee
so that the boundary conditions for ${\cal A}$ at the boundary
$\partial M$ of $M$ are expressed through the generating parameters
$A_\alpha$.
One of the names to the average at the l.h.s. of (\ref{basrel})
is (generalized) $\tau$-function and (perhaps for somewhat
sophisticated groups) it can be considered as providing an
effective action of (actually, {\it any}) quantum field theory,
while $\{A_\alpha\}$ can be considered as some particular choice
of coupling constants (comp. with (\ref{branetrue})).
Then relation (\ref{basrel}) implies that
the $\tau$-function at the l.h.s. can be represented
as a functional integral over gauge fields in some gauge
topological field theory in non-compact space time.
In other words, there is supposedly
a mapping from the space of $d$-dimensional
QFT models into that of topological $d+1$-dimensional models
and it is established through the study of integrable
structure of effective actions.
Moreover, according to (\ref{basrel}), integrable properties
should be exhibited not only in effective action's
dependence on the coupling constants but also in that on
the boundary conditions (the choice of vacuum).
This suggestion can provide a new powerful tool for the study
of (generalized) $\tau$-functions, which -- if attacked
straightforwadly -- is a difficult task already for 2-loop algebras.

{\bf 9.} Let us consider relation (\ref{basrel}) for the case of
affine algebra ${\cal G} = \hat{U(1)}_{k=1}$
with unit central charge $k=1$. Then the average at the l.h.s.
can be represented as a correlator in the theory of free
fermions on a Riemann surface ${\cal C}$,
\be
\langle \ldots \rangle \ =
\int {\cal D}\tilde\psi {\cal D}\psi\ (\ldots)\ \exp
\int_{\cal C} \tilde\psi \bar\partial \psi
\ee
and the generators  $J$ constitute holomorphic current
$J(z) = \tilde\psi\psi(z)$.
As explained above, the
topological theory at the r.h.s. of (\ref{basrel}) in this
case is abelian Chern-Simons model on the $d=3$ manifold $M$,
obtained by ``filling'' the Riemann surface ${\cal C} = \partial M$,
and (\ref{basrel}) in this case is nothing but (\ref{KK2}) .
Boundary conditions at ${\cal C}$ can be imposed only on one
of components of the gauge field ${\cal A}$, the others remain
unconstrained. The relevant choice is the antiholomorphic
component $\bar A(z,\bar z)$ ($z$ are coordinates on ${\cal C}$).

Finally, (\ref{basrel}) for affine algebra $\hat{U(1)}$ with
the central charge $k=1$ acquires the form:

\be
\tau\{\bar A|{\cal C}\} =\
\langle \exp \int_{{\cal C}=\partial M} \bar A J \rangle\ =
\nn \\ =
\int_{\left.{\cal A}\right|_{\partial M = {\cal C}} = \bar A}
{\cal DA} \exp \left(\int_M {\cal A}d{\cal A} +
\oint_{\partial M} A\bar A\right)
\label{basrelU1}
\ee
The most straightforward proof of this relation is just
independent  calculation of both sides, which gives:\footnote{
On the fermionic side one should use conservation of the current,
$\bar\partial J(z) = 0$ to express surface integral through
contour integrals, transform exponent to the normal form and
make use of the expression
$$
\langle :\exp\left( \delta_i\oint_{A_i} J +
\epsilon_i\oint_{B_i} J\right): \rangle =
\theta_{\vec e}\left(\vec{\epsilon T} + \vec\delta\right)
$$}

\be
\tau\{\bar A|{\cal C}\} =\
\theta_{\vec e}\left(\int_{\cal C}\bar A\vec \omega\right)
\cdot \exp \int_{\cal C}\int_{\cal C}  \bar A(x)
G(x,y)\bar A(y)
\ee
with the Green function \cite{Fay,freef,rgw}
\be
G(x,y) = \Psi_{\vec a}(x,y)\Psi_{-\vec a}(x,y)  = \nn \\
= \partial_x\partial_y\log E(x,y) + \sum_{i,j}\omega_i\omega_j
\partial^2_{ij}\log\theta_{\vec a}(\vec x - \vec y)
\ee
expressed through Szogo kernel
\be
\Psi_{\vec a}(x,y) = \frac{\theta_*(\vec x - \vec y + \vec a)}
{\theta_*(\vec a) E(x,y)} =
\frac{\theta_{\vec a}(\vec x - \vec y)}
{\theta_{\vec a}(\vec 0) E(x,y)}
\ee
The characteristic
\be
\vec a = \vec e - \vec e_* + \int_{\cal C}\bar A\vec \omega
\ee
is expressed through the even half-integer theta-characteristic
${\vec e}$, used to define the free-fermion model on ${\cal C}$,
some odd half-integer theta-characteristic $\vec e_*$ used to define
the theta-function $\theta_*$,
$\theta_{\vec e} = \theta_*(\vec e - \vec e_*)$,
and the integral over ${\cal C}$
of the product of $\bar A(z,\bar z)$ times holomorphic
$(1,0)$-differentials $\vec \omega(z)$.

The l.h.s. of (\ref{basrelU1}) is nothing but a $\tau$-function
of KP/Toda-family. Conventional Krichever's $\tau$-function
\cite{kridata} appears when
\be
\bar A = \bar\partial \left(\sum_{k=1^\infty}
\frac{t_k}{(z-z_0)^k}\right) =
\sum_{k=1}^\infty \frac{(-)^kt_k}{(k-1)!}
\partial^{k-1}\delta^{(2)}(z-z_0)
\ee
so that
\be
\langle \exp \int_{{\cal C}} \bar A J \rangle\ =
\ \langle \exp \sum_{k=1}^\infty t_k J^{(z_0)}_k \rangle
\ee
As usual, in addition to the time-variables $t_k$,
it depends on the set of Krichever data \cite{kridata}:
a Riemann surface ${\cal C}$, a point $z_0$ on it
and coordinates $z$ in the vicinity of $z_0$.
Given this data, one can define
$J(z) = \sum_k J_k^{(z_0)} (z-z_0)^{k-1} dz$.
If $supp(\bar A)$ consists of $n$ points
$z_{01},\ldots,z_{0n}$,

\be
\bar A = \bar\partial \sum_{i=1}^n\left(\sum_{k=1^\infty}
\frac{t_k^{(i)}}{(z-z_{0i})^k}\right)
\ee
we get the so-called $n$-component  KP/Toda $\tau$-function
($n=1$ is the KP and $n=2$ the ordinary ``Toda-lattice'' case).
In this sense we have at the l.h.s. of (\ref{basrelU1})
a {\it generic} ($\infty$-component) KP/Toda $\tau$-function.
As every (generalized) $\tau$-function, it
satisfies bilinear Hirota equation \cite{GKLMM}.
The Miwa transform, which produces an insertion of
$\psi(\lambda)$ or $\tilde\psi(\lambda)$ in the fermionic
correlator, is just a shift
\be
\bar A \rightarrow \bar A + \frac{1}{\bar z - \bar\lambda}
\ee

Thus -- as a simple example of ``AdS/CFT correspondence'' --
we see that partition function of Chern-Simons theory
on a manifold with a boundary is -- as a function of
the boundary conditions -- a $\tau$-function of the
KP/Toda family (i.e. associated with the theory of free
fermions on Riemann surfaces).
This example reveals -- in a simple situation --
integrable properties of effective actions as
functionals of the boundary conditions (the moduli of
vacua manifolds). This supplements
the usual conjecture \cite{tom, ufn2} about integrable
dependence of effective actions on the coupling
constants.

{\bf 10.} Another relation between the ``AdS/CFT correspondence''
and group theory is provided by representation theory of
non-compact groups. The most peculiar phenomenon here is
the occurrence of ``singletons'', which can be considered
as fundamental representations (i.e. all other relevant
representations belong to some power of the fundamental
one) and at the same time can be interpreted as localized
at the boundary of the non-compact homogeneous space
where the group is acting. There are plenty of interesting
speculations relating the singleton-like phenomena to
Kaluza-Klein theories \cite{singl} though not all the
sides of the story are fully clarified. The story is
closely related to the theory of Harish-Chandra functions
and can again be attacked by the methods of geometrical
quantization \cite{HC}.

The simplest case where the phenomenon is already present
(at least if somewhat obscure notion of localization at
the boundary is substituted by the leading asymptotics
at the boundary) is $AdS_2$: the homogeneous space of
the $d=2$ conformal group $SO(d-1,2) =
SO(1,2) \cong SL(2,R)$, i.e. the Lobachevsky plane.
Lobachevsky plane can be represented as an upper half-plane $y>0$,
of the complex plane with coordinate $z= x+iy$,
metric $ds^2 = \frac{dz d\bar z}{y^2}$ and
Laplace operator:
\be
\Delta = y^2\partial\bar\partial
\ee
Eigenfunctions of Laplace operator are \cite{Gel}
\be
f_p(x,y) = \int da f(a) \left(\frac{(az+1)(a\bar z+ 1)}{y}\right)^p
\ee
with any function $f(a)$. The corresponding eigenvalue is
$\lambda = p(p-1)$.  Representation is unitary if
$p = -\frac{1}{2} + in$ with integer $n$.
Since
\be
f_p(x,y) = y^{-p} \int da f(a) (ax + 1)^{2p} \left( 1 + O(y)\right)
\ee
the leading asymptotics at the boundary $y=0$ of the entire set
of eigenfunctions with the given eigenvalue is just the set $F_{2p}$
of homogeneous functions of weight $2p$ (under rational
transformations $z \rightarrow \frac{az +b}{cz+d}$).
For example, for integer $p$ this set is just the linear space of
all the polynomials (in $x$) of degree $2p$.
Remarkably (and obviously) for integer $2p$
\be
F_{2p} = F_2^{\otimes p}
\ee

{\bf 11.} I am indebted for numerous discussions on the subjects
of above notes to
E.Akh\-medov, L.Dolan, A.Gera\-simov, A.Gorsky, S.Gukov, A.Losev,
A.Mar\-shakov, A.Mi\-khai\-lov, A.Mironov and A.Rosly.

It is my pleasure to aknowledge the hospitality of Centro A.Volta
and the support of INTAS-RFBR to this School through the grant
95-1353.
My work is partly supported by the Russian President's grant
96-15-96939 and RFBR grant 98-02-16575.

\end{document}